\documentclass[%
 aip,
 amsmath,amssymb,
 reprint,%
]{revtex4-1}

\usepackage{graphicx}% Include figure files
\usepackage{dcolumn}% Align table columns on decimal point
\usepackage{bm}% bold math
\usepackage[utf8]{inputenc}
\usepackage[T1]{fontenc}
\usepackage{mathptmx}
\usepackage[capitalise]{cleveref}
\usepackage{chemformula}

\begin{document}

\preprint{AIP/123-QED}

%\title[Sample title]{Sample Title:\\with Forced Linebreak}
\title{Spin-state smectics in spin crossover materials}
% Force line breaks with \\

\author{J. Cruddas}
\author{G. Ruzzi}%
\affiliation{ 
School of Mathematics and Physics, The University of Queensland, Brisbane, Queensland 4071, Australia
}%

\author{B. J. Powell}
\email{powell@physics.uq.edu.au}
\affiliation{ 
	School of Mathematics and Physics, The University of Queensland, Brisbane, Queensland 4071, Australia
}%

\date{\today}% It is always \today, today,

\begin{abstract}
We show that a simple two dimensional model of spin crossover materials gives rise to spin-state smectic phases where the pattern of high-spin (HS) and low-spin (LS) metal centers spontaneously breaks rotational symmetry and translational symmetry in one direction only. The spin-state smectics are distinct thermodynamic phases and give rise to plateaus in the fraction of HS metal centers. Smectic order leads to lines of Bragg peaks in the x-ray and neutron scattering structure factors.   We identify two smectic phases and show that both are ordered in one direction, but disordered in the other, and hence that their residual entropy scales with the linear dimension of the system. This is intermediate to spin-state ices (examples of `spin-state liquids') where the residual entropy scales with the system volume, and antiferroeleastic ordered phases (examples of `spin-state crystals') where the residual entropy is independent of the size of the system. 
\end{abstract}

\maketitle

\section{Introduction}

Spontaneous symmetry breaking: the emergence, at low-temperatures, of long-range order from a high-temperature disordered state is one of the foundations of condensed matter physics.\cite{AndersonBasic,Contempt} For example in crystals translational and rotational symmetries, that are present in liquids and gases, are spontaneously broken. This has profound consequences: for example it leads to the rigidity of crystals and predicts the existance of massless (gapless, linearly dispersing) low-energy excitations (acoustic phonons).

However, it has become increasingly clear that when disordered states survive to low temperatures in strongly interacting systems it is often a sign that something extremely interesting is happening. Key examples include the Tomonaga-Luttinger liquid in one-dimension and quantum spin liquids in two and three dimensions.\cite{Contempt,Savary,Giamachi} Here quantum fluctuations are sufficiently strong to entirely suppress long-range order. But classical physics can also lead to low-temperature disordered states.\cite{Goodwin} In classical systems, entropy rather than quantum fluctuations, is responsible for the stabilization of a disordered state. For example, in ice-states there are a macroscopically large number of microstates with the same energy. Thus, there is a large residual (zero temperature) entropy and the system remains disordered a low temperatures. The two best known examples are proton disorder in I$_h$ and I$_c$ water ice and the magnetic disorder in spin ices,\cite{Pauling,CastelnovoARCMP,Henley} but recently ice phases have been discovered or predicted in many other systems,\cite{Goodwin}
including spin crossover (SCO) materials.\cite{JaceKagome,JacePyro}

Disorder does not imply that the system is completely random. In gas-like states randomness emerges from the weakness of the interactions between the constituents. But, in liquid-like states the low-energy physics is dictated by the strong correlations between the constituents.\cite{SciP} The classic example is the Bernal-Fowler ice rules in water ice, which dictate that locally every oxygen forms covalent bonds with exactly two protons.\cite{Bernal} There are an extremely large number of microstates consistent with the ice rules meaning that the instantaneous configurations look random on large length-scales. Nevertheless the behavior of each proton  is stongly correlated with those of other protons nearby. This, and similar ice rules in other systems, result in distinctive signatures in diffraction experiments.\cite{Henley,Fennell,Morris}

%However, ice-physics is not the only route to low-temperature disordered phases in classical systems.
%Correlated disordered phases are crucial for the (potential) uses of many materials including photovoltaics (hybrid organic-inorganic perovskites), ferroelectrics (e.g., \ch{BaTiO3}), and magnets (spinels).\cite{Goodwin} Further applications are being actively investigated, therefore a key goal is to identify new routes to disordered materials that provide resources for future potential applications.

Liquid crystals host phases of matter intermediate between crystalline and liquid phases -- rotational order is spontaneously broken in both nematic and smectic phases, and translational symmetry is also spontaneously broken in one direction (but not in the perpendicular directions) of smectic phases.\cite{chaikin2000principles}  Thus crystals are more ordered than smectics, smectics  are more ordered than nematics, and nematics are more ordered than liquids. This endows liquid crystals with properties that have proved incredibly technologically useful.\cite{Castellano} Liquid-crystal-like phases -- where rotational symmetry is broken, but translational symmetry is (partially) preserved -- have been identified in other systems, including electronic nematics,\cite{Schmalian,Kivelson1998,Kuo958,Chu710} electronic smectics,\cite{Emery,Yim2018} spin nematics,\cite{Penc2011} and spin smectics.\cite{Lebert20280}
 
\begin{figure*}
	\includegraphics[width=0.825\linewidth]{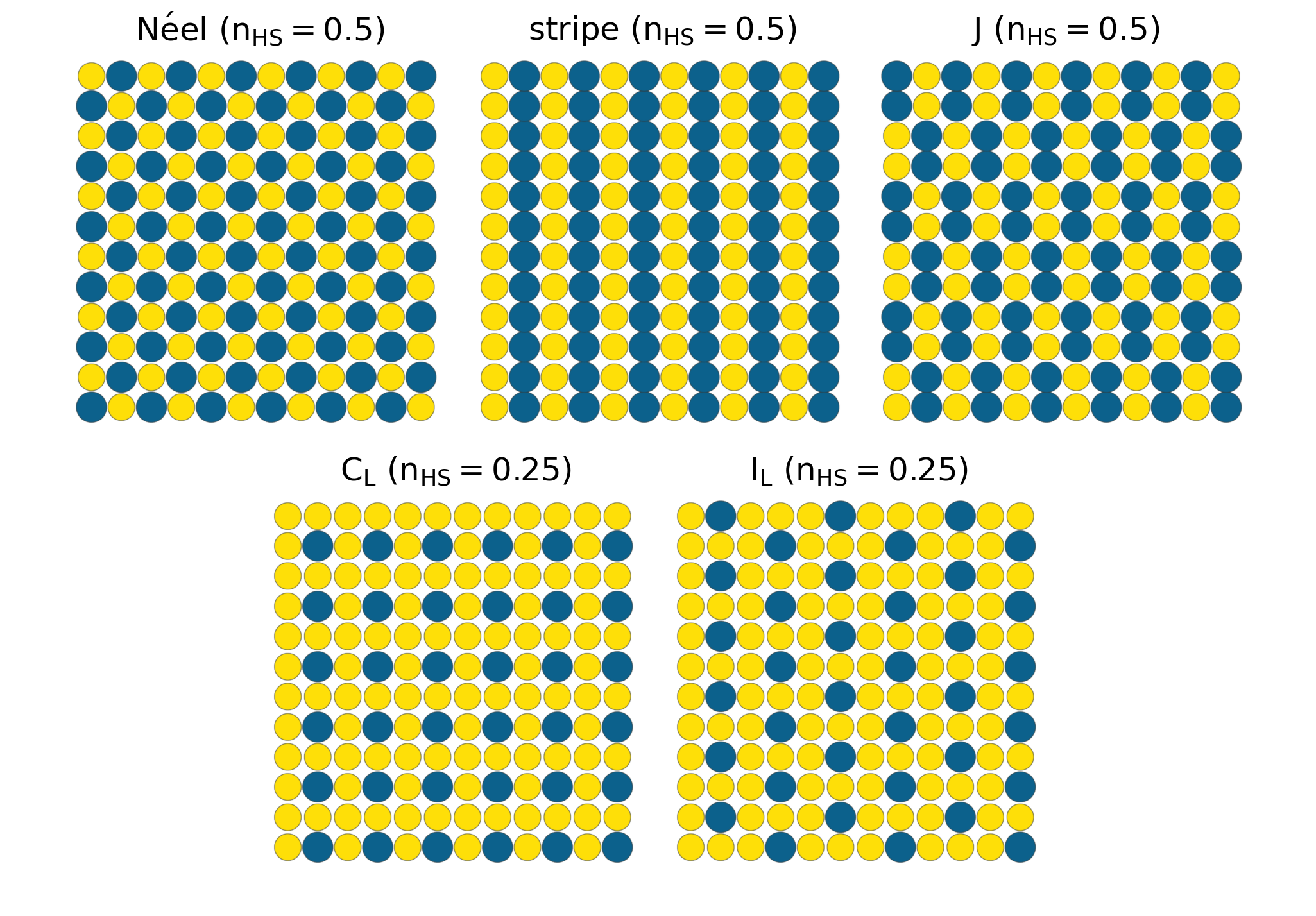}
	\caption{\label{fig:sketch} The selected spin-state crystalline orderings, with long-range order in both directions, observed in our Monte Carlo simulations. The LS (HS) sites are indicated by blue (yellow) circles. For clarity, we have retained the labels used in our previous work\cite{JaceSquare}. Where $n_{HS}\ne1/2$ we show only the majority LS variant, indicated by a subscript L, interchanging the spin state of all molecules results in a majority HS variant, indicated by a subscript H.}
\end{figure*}

Here we demonstrate that smectic phases  occur in a two-dimensional model of SCO materials. We show that the interplay between competing crystalline orders and the entropy of mixing can lead to phases, in which rotational symmetry is broken and translational symmetry is broken in one direction, but preserved in the perpendicular direction. Thus, spin-state smectics are intermediate between spin-state liquids (of which spin-state ice is the only concrete example proposed so far) and spin-state crystals (where the pattern of spin-states spontaneously breaks translations symmetry in all directions, see \cref{fig:sketch} for some examples). We show that this hierarchy can be quantified by the scaling behavior of the residual entropy with the system size.

We show that spin-state smectic phases lead to plateaus in the fraction of high-spin states, $n_{HS}$ (which is experimentally measurable via $\chi T$, where $\chi$ is the magnetic susceptibility and $T$ is the temperature). There are several materials where plateaus in $\chi T$ have been observed, but no long-range order in the spin-states has been resolved.\cite{sciortino2017,crevasse,OrderDisorder} Smectic phases offer a possible explanation for these experiments. We discuss the signatures of smectic phases in x-ray and neutron scattering experiments, which would allow for a definitive identification of a smectic phase. 

\section{Model and methods}

\begin{figure}
	\includegraphics[width=0.495\columnwidth]{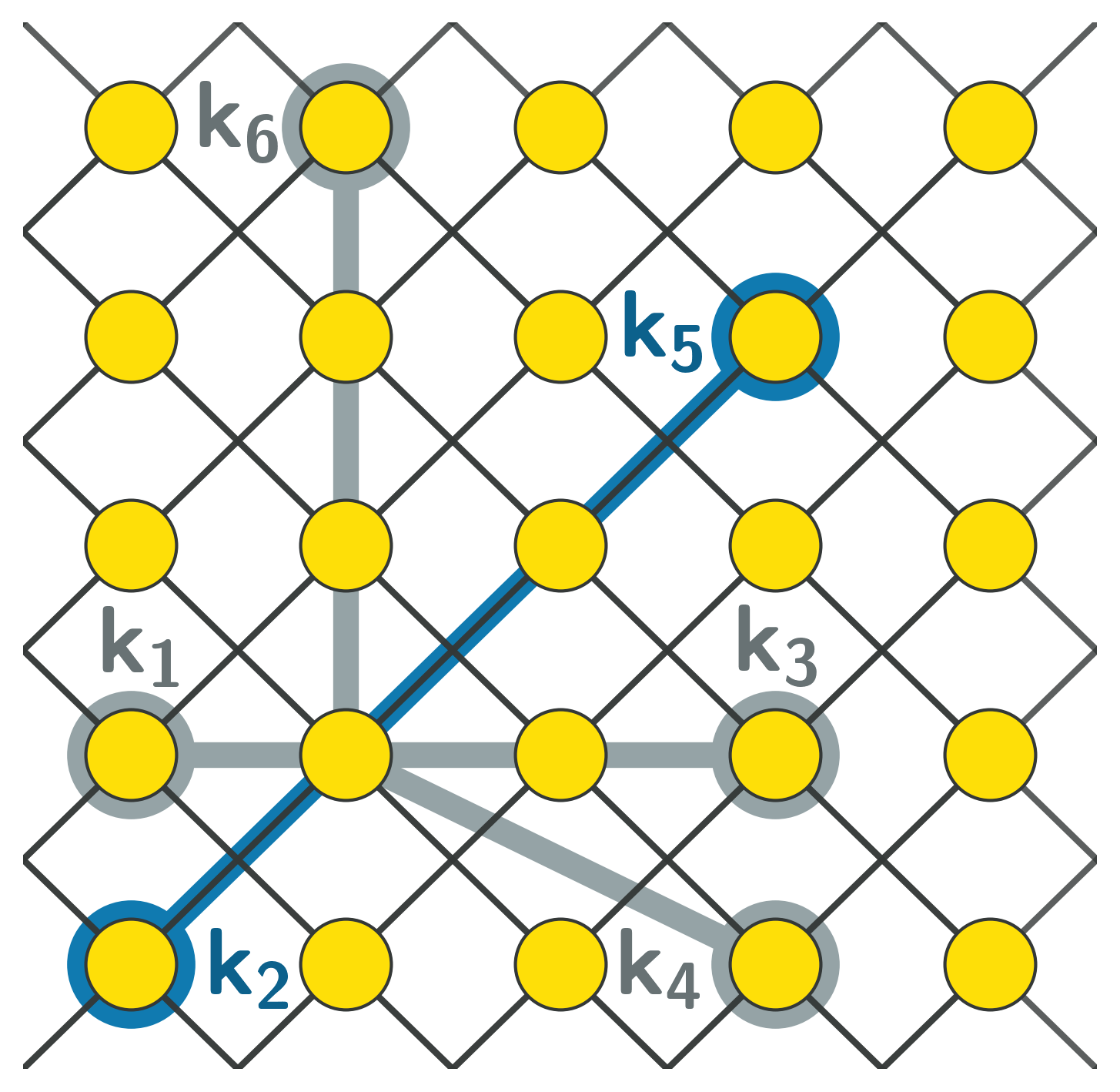}
	\includegraphics[width=0.495\columnwidth]{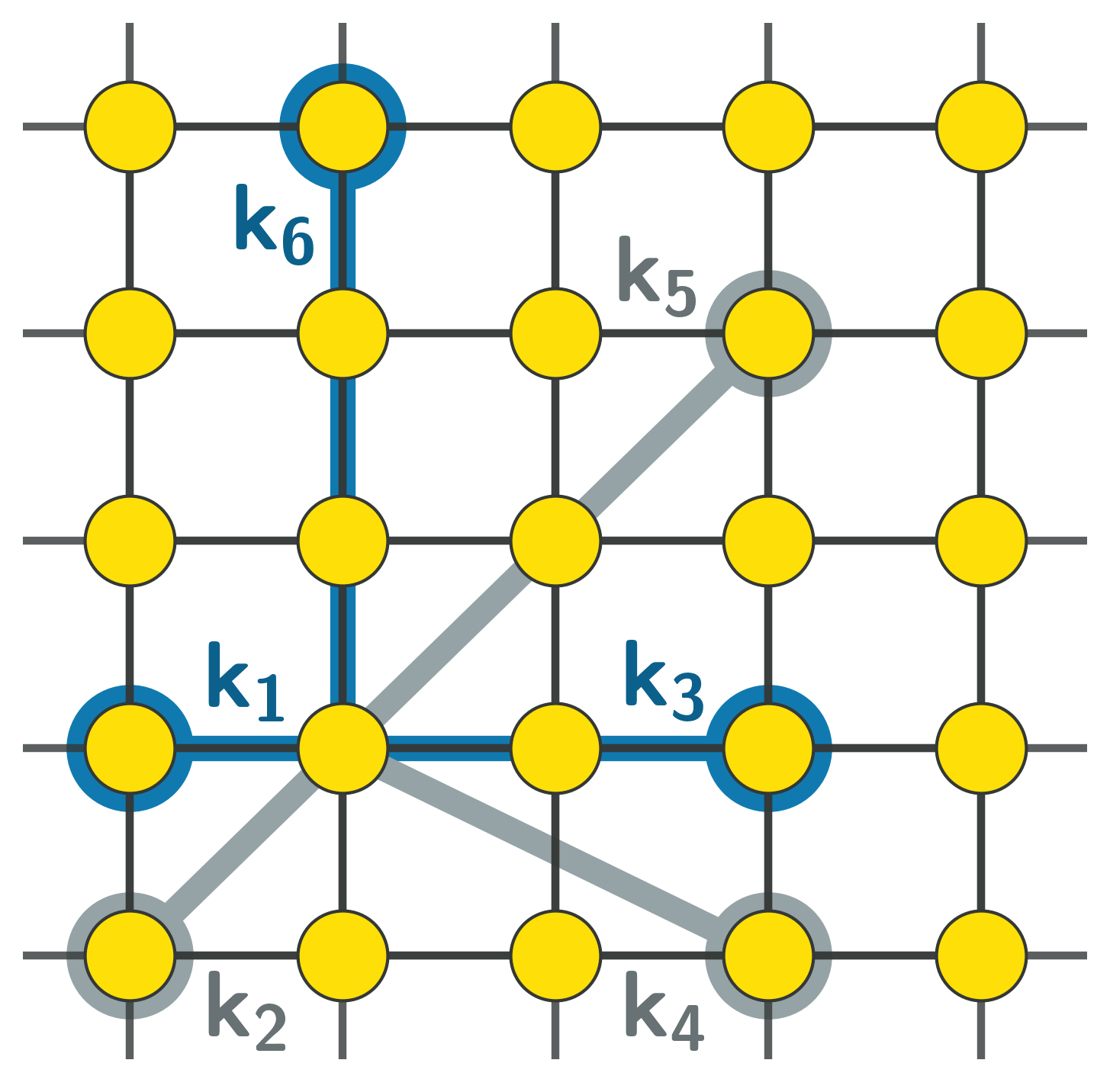}% Here is how to import EPS art
	\caption{\label{fig:square} Examples of Hofmann frameworks topologically equivalent to the square lattice. The metal centres are marked by yellow circles and in-plane ligands are marked by black lines. The elastic interactions between the metal centres, $k_n$, are marked for the $n$th nearest neighbours, where $n\in \{1,2 ,\dots,6\}$. The spring constants ($k_n$) are expected to be typically positive for through-bond interactions (blue lines and circles) and typically negative for through-space interactions (grey lines and circles) \cite{JaceSquare}.}
\end{figure}

We consider a square lattice of SCO molecules coupled by springs (\cref{fig:square}) and described by the Hamiltonian
\begin{eqnarray}
	\mathcal{H} &=& \frac{1}{2}\sum_i \left( \Delta H - T \Delta S \right) \notag\\&&
	+ \sum_{n=1}^5 \frac{k_n}{2} \sum_{\langle i,j\rangle_n} \left\{	 r_{i,j} - \eta_n \left[ \overline{R} + \delta(\sigma_i+\sigma_j) \right] \right\},
	\label{Hsprings}
\end{eqnarray}
where $\Delta H=H^{HS} - H^{LS}$ is the enthalpy difference between the HS and LS states of an individual molecule,\cite{Miriam}  $\Delta S=S^{HS} - S^{LS}$ is the entropy distance between the HS and LS states of an individual molecule, $k_n$ is the spring constant between $n$th nearest neighbors, $r_{i,j}$ is the instantaneous difference between the $i$th and $j$th molecules, $\eta_n=1,\, \sqrt{2}, 2,\, \sqrt{5},\, 2\sqrt{2},\, \dots$ is the ratio of distances between the $n$th and 1st nearest-neighbor distance on the undistorted square lattice, $\overline{R}=(R_{HS}+R_{LS})/2$, $\delta=(R_{HS}-R_{LS})/4$, $R_{HS}$ ($R_{LS}$) is the average distance between the centers of nearest neighbor molecules in the HS (LS) phases, and the pseudospin degrees of freedom are $\sigma_i=1$ ($-1$) if the i$th$ molecule is HS (LS). This, and closely related models, have been widely studied previously and provided many insights into the physics of SCO materials.\cite{pavlik2013,Gutlich2000,NishinoElastic2009,Nishino2007,Nishino15,Nishino19,Miyashita,Nakada2011,Nakada2012,Nishino13,Watanabe,EnachescuMC2010,EnachescuMC2012,EnachescuMCDiluted2011,Enachescu17,Paez2016,KonishiMC2008,Ye2015,Stoleriu,JaceKagome,JaceSquare,JacePyro,GianExact}

We make the symmetric breathing mode approximation (SBMA),\cite{JaceKagome,JaceSquare} that is, we assume that for all nearest neighbors,
$r_{i,j} = x$, and that the topology of the lattice is not altered by the changes in the spin-states. In this approximation the Hamiltonian (\ref{Hsprings}) becomes an Ising-Husimi-Temperley model in a longitudinal field:
\begin{equation}
\mathcal{H}
\approx\sum_{n=1}^5 J_n\sum_{\langle i,j \rangle_n}\sigma_i\sigma_j-\frac{J_\infty}{N}\sum_{i,j}\sigma_i\sigma_j
+\frac{1}{2}\sum_i \Delta G_i\sigma_i,\label{model}
\end{equation}
where, $J_n=k_n \eta_n^2 \delta^2$ is the pseudospin-pseudospin interaction between  the $n$th nearest-neighbors,
$J_\infty=\delta^2 \sum_{n=1}^m(k_n z_n \eta_n^2)$ is the long-range strain, 
$\Delta G=\Delta G^{HS}-\Delta G^{LS}=\Delta H - T \Delta S$ is the free energy difference between the HS and LS states of the $i$th molecule,
$z_n$ is  the coordination number for \textit{n}th nearest neighbors, and 
$N$ is the number of  sites. 
%The most important artifact of the SBMA is that it overestimates the strain -- the SBMA gives an infinite range strain that is constant at large distances, whereas an exact treatment\cite{GianExact} would give a merely long-ranged strain, decaying with a power law. As the strain is ferroelastic $J_{\infty}>0$ this always the HS and LS phases over antiferroelastic order or disordered phases. Therefore, the SMBA always suppresses the emergent behaviors described below.

The spring constants should be typically positive for through-bond interactions, but will often be negative for through space interactions. \cite{JaceSquare} This has profound consequences for the long-range order observed in different materials. However, the possible range of spring constants is constrained by the fact that the lattice described by Hamiltonian (\ref{Hsprings}) must be stable, i.e., we must have $\partial^2\mathcal{H}/\partial x^2\propto J_{\infty}>0$ or, equivalently,  $\sum_{n}k_n z_n \eta_n^2>0$.

We solve the Ising-Husimi-Temperley model (Eq. (\ref{model})) quasianalytically at $T=0$. That is, we consider all possible states with a unit cell no larger than $4 \times 4$ sites.  The $T=0$ phase is then set to be the state with the lowest energy. For $T>0$ we solve the model via Monte Carlo simulations on a $N=60\times60$ lattice with periodic boundary conditions. 

We perform three types of Monte Carlo simulations: heating, cooling and parallel tempering. For cooling (heating) runs we initialize the calculation in a random configuration (the $T=0$ ground state) and lower (raise) the temperature in steps of $1000$ Monte Carlo steps, retaining the final configuration of the last step as the first configuration of the new step. Parallel tempering runs employ $300$ copies of the simulation initialized in a random configuration and allow accurate determination of the lowest free energy state.  

For each heating, cooling and parallel tempering run we use single spin flip, loop and worm algorithms.\cite{Newman} The loop algorithm works by creating a Monte Carlo update that maps between two states with the same local correlations. For the smectic phases in this paper, the loop update is always a line of flipped spins. The worm algorithm works by creating a Monte Carlo update that removes two defects (violations of the local correlations) of the opposite polarization by annihilating them with one another. 

The spin flip, loop, and worm algorithms are integrated into a single update. We  choose a single spin flip update with probability $1-1/N$ and  a combined loop/worm update with probability $1/N$. The latter  picks a chain of neighboring sites and calculates the enthalpy change if the spin-states of all of these sites are changed, starting with a pair of sites and expanding from the end of the chain.  This process terminates when either an energetically favorable update is generated or if the Boltzmann probability for the move is less than a randomly selected number, or if the chain terminates on itself creating a chain and a loop. The Boltzmann probability of the loop update occurring is then checked.

The spin-state structure factor is given by 
\begin{equation}
S(\bm q) = \frac{1}{N^2} \sum_{i,j} \langle \sigma_i \sigma_j \rangle e^{-i\bm q\cdot(\bm r_i-\bm r_j)},
\label{eq:SF}
\end{equation}
where $\bm r_i$ is the position of the $i$th site.
The average was evaluated over $100$ configurations, each separated by $N$ Monte Carlo steps, during a parallel tempering calculation.

\section{Results}

\subsection{Disordered phases due to competing orders}

We have previously shown that a wide range of long-range antiferroelastic ordered phases can occur in an Ising-Husimi-Temperley model of SCO materials, \cref{model}.\cite{JaceSquare} These phases show long-range order in both directions, which we will henceforth refer to as spin-state crystals.\cite{SciP} In this paper we will demonstrate that smectic phases, which are ordered in one spatial direction but disordered in the perpendicular direction, are also natural and that, at non-zero temperatures -- where all experiments are carried out,  achieving these phases does not require fine tuning.

\begin{figure}
	\includegraphics[width=\columnwidth]{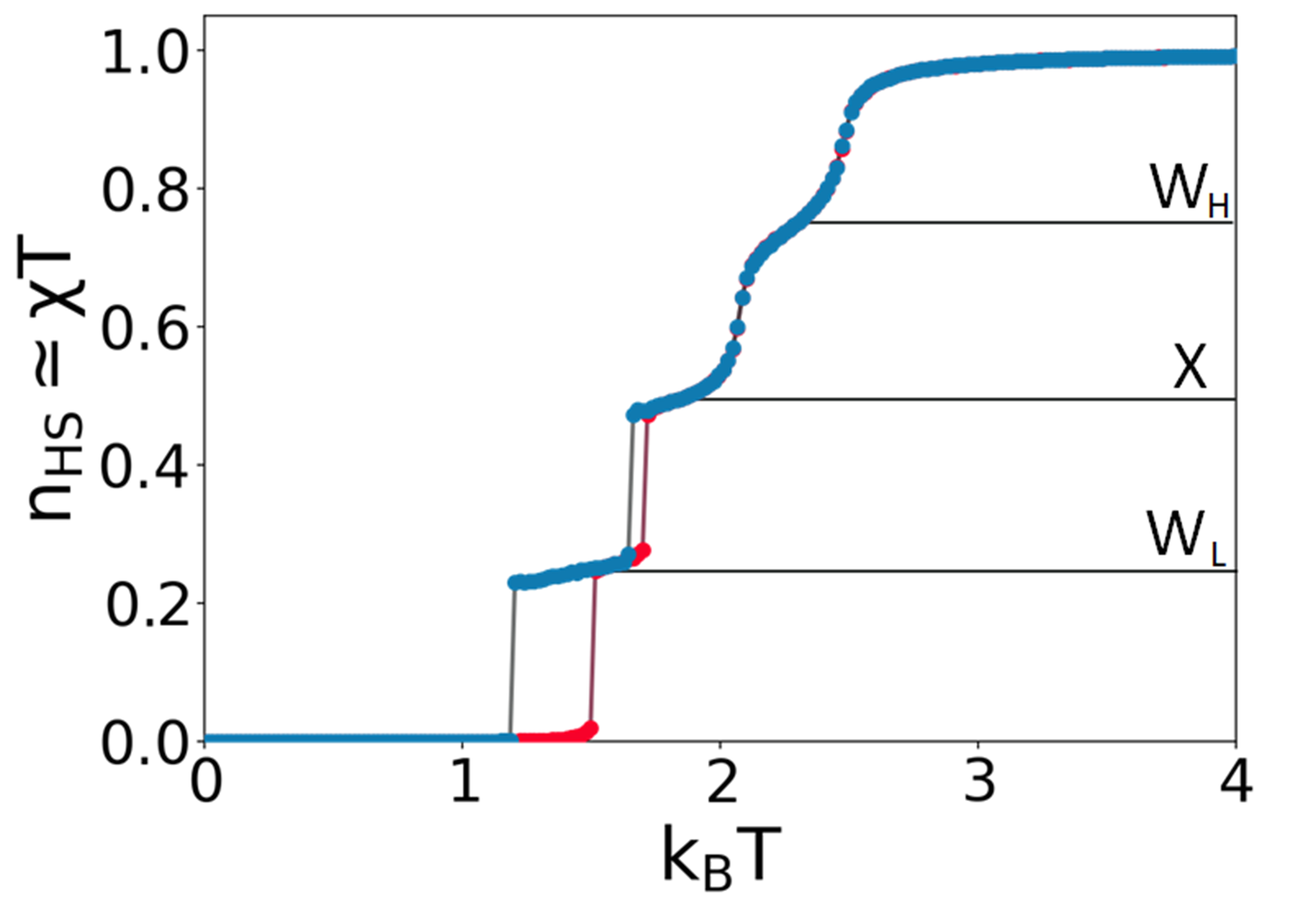}% Here is how to import EPS art
	\caption{\label{fig:nHS} An example of the calculated fraction of high spins, $n_{HS}$, in a four-step transition. Results are for $\Delta S=4\log 5$, $\Delta H=12.1k_1\delta^2$, $k_1>0$, $k_2=k_1/4$, $k_3=-0.39 k_1$, $k_4=0$, and $k_5=0.0975k_1$. The red and blue lines indicate heating and cooling respectively. }
\end{figure}

\begin{figure}	
		\includegraphics[width=0.9\columnwidth]{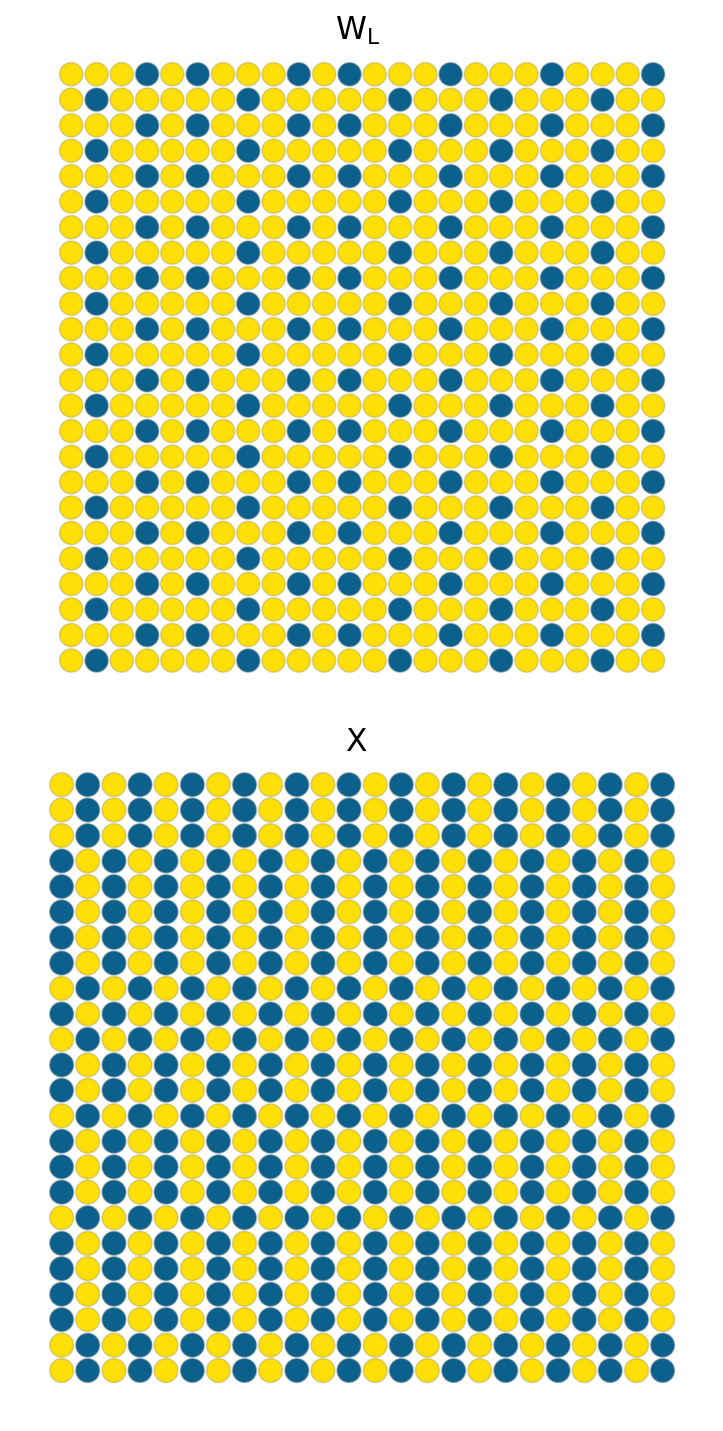}
	\caption{\label{fig:snapshots} Truncated snapshots of the spin-state smectic W$_\text{L}$ and X phases from our Monte Carlo simulations.  LS (HS) sites are indicated by blue (yellow) circles. }
\end{figure}

An example of a  four-step SCO transition with intermediate plateaus at $n_{HS}=0.75$, $0.5$ and $0.25$ is shown in \cref{fig:nHS}.  We have previously reported several examples of similar transitions in this model,\cite{JaceSquare} in those cases the plateaus are associated with spin-state crystals. However, when we plot snapshots, \cref{fig:snapshots}, of the  Monte Carlo calculations reported in \cref{fig:nHS} we find long-range order in only one direction in the intermediate plateaus (labeled  W$_\text{H}$, W$_\text{L}$, and X). This is highly reminiscent of smectic phases in liquid crystals.

%Rather than long-range order, in the  W$_\text{H}$, W$_\text{L}$, and X plateaus we observe only short-ranged correlations (the correlations are much smaller length-scale than the domain size one would usually expect for  a long-range ordered phase). The W plateaus display short-range correlations characteristic of the C and I phases (sketched in \cref{fig:sketch}) and the X plateau displays short-range correlations characteristic of the N\'eel (also known as checkerboard), stripe, and J phases (\cref{fig:sketch}). Note that the N\'eel, stripe, and J phases all have $n_{HS}=1/2$, which is consistent with the observation the $n_{HS}\simeq0.5$ in the  X plateau, and the C$_\text{H}$ and I$_\text{H}$ (C$_\text{L}$ and I$_\text{L}$) phases have $n_{HS}=0.75$ ($n_{HS}=0.25$) consistent with the calculated fraction of HSs in the W$_\text{H}$ (W$_\text{L}$) plateau.

\begin{figure}
	\includegraphics[width=\columnwidth]{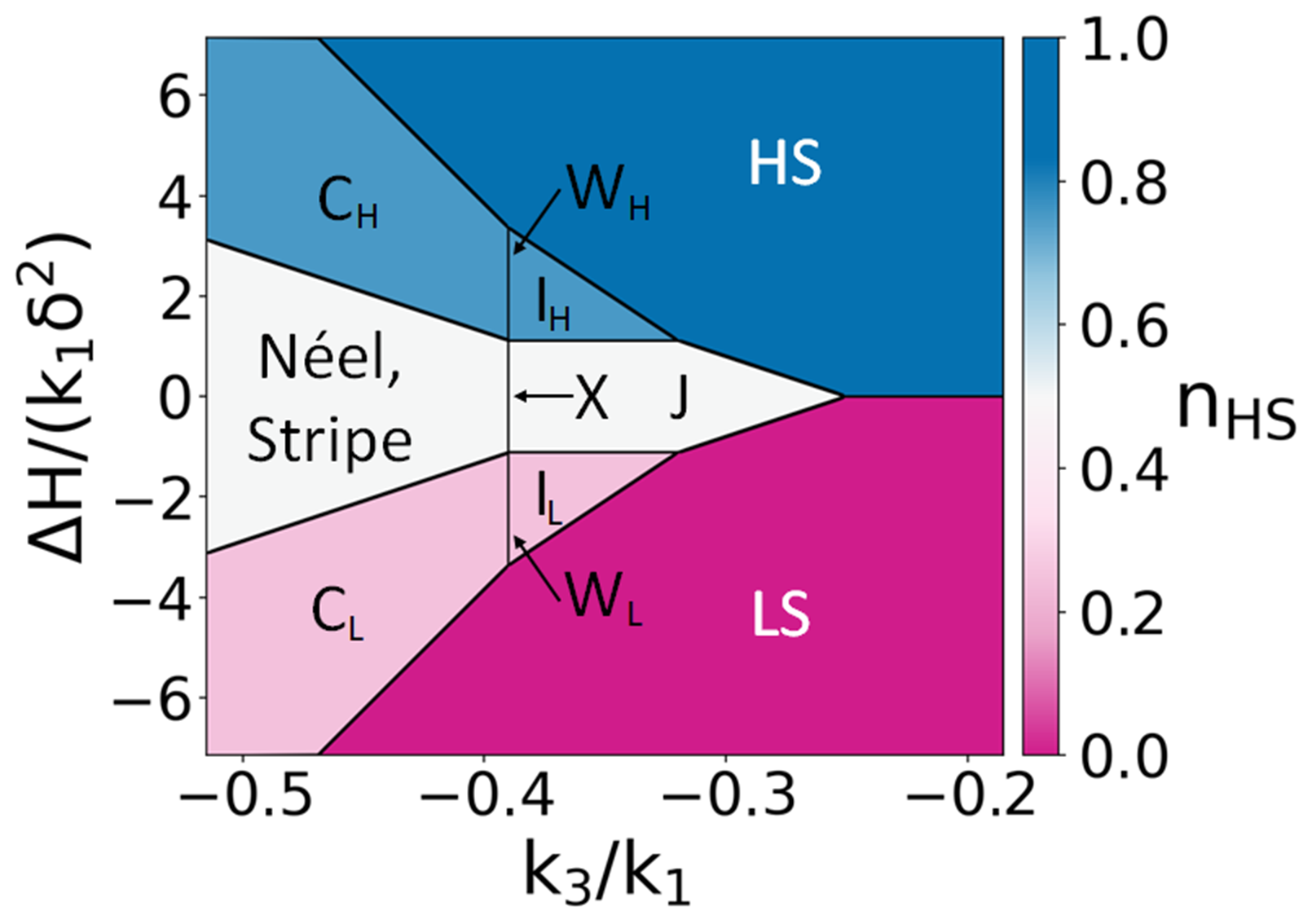}
	\caption{\label{fig:phaseDia0T} A slice of the $T=0$  phase diagram.  
		Shading represents the fraction of high spins, $n_{HS}$. The solid black lines are first order transitions.
		Here $k_1>0$, $k_2=k_1/4$, $k_4=0$, and $k_5=0.0975k_1$; 	
	}
\end{figure}

To better understand the origin of these smectic phases we show a slice of the $T=0$ phase diagram for the model  in \cref{fig:phaseDia0T}. A wide variety of  spin-state crystals are found in this model\cite{JaceSquare}; those discussed in this paper are shown in \cref{fig:sketch}. 
Note that for the parameters studied in \cref{fig:phaseDia0T} the N\'eel and stripe phases are degenerate (have the same free energy). This is an accidental degeneracy  due to the parameters chosen. This fine tuning is not necessary for the physics described below, rather we have chosen to plot this slice of the (eight-dimensional) phase diagram as it simplifies the discussion below.
In the regions marked ``N\'eel, stripe''  in \cref{fig:phaseDia0T} we find that either the N\'eel or the stripe phase is the lowest energy state; since the domain walls between the N\'eel and  stripe phases  carry a non-zero energy cost coexistence does not occur at $T=0$. However, in Monte Carlo calculations at finite temperatures, on cooling into these regions of the phase diagram, we sometimes find domains of both phases. %, however these domains are much larger than those found in the disordered  W$_\text{H}$, W$_\text{L}$, and X plateaus in Figs. \ref{fig:nHS} and \ref{fig:snapshots}. 

At $T=0$ the spin state crystal phases are separated by first order transitions, marked by solid lines (second order transitions are not possible at $T=0$ in classical models). Theoretically, we can fine-tune the parameters to examine the properties of the model exactly at the phase transition: for example at the points marked X (on the boundary between the N\'eel, stripe and J phases) or W  (on the boundaries between the C and I phases) in \cref{fig:phaseDia0T}. By definition, at the phase boundary the energies of the two competing phases are equal. Therefore, the energetic cost of forming a mixture is set only by the energy required to form domain walls. It has been demonstrated previously that, in the Ising model at $T=0$, if one fine tunes to certain points along the phase boundary, then one can find a disordered state with a residual entropy.\cite{Kassan-Ogly}
Experimentally such fine tuning would be difficult to achieve in SCO materials. Nevertheless, we will see below that at non-zero temperatures the entropy of mixing becomes important and the single point is expanded to a phase spanning a broad region in parameter space.

\begin{figure}
	\includegraphics[width=\columnwidth]{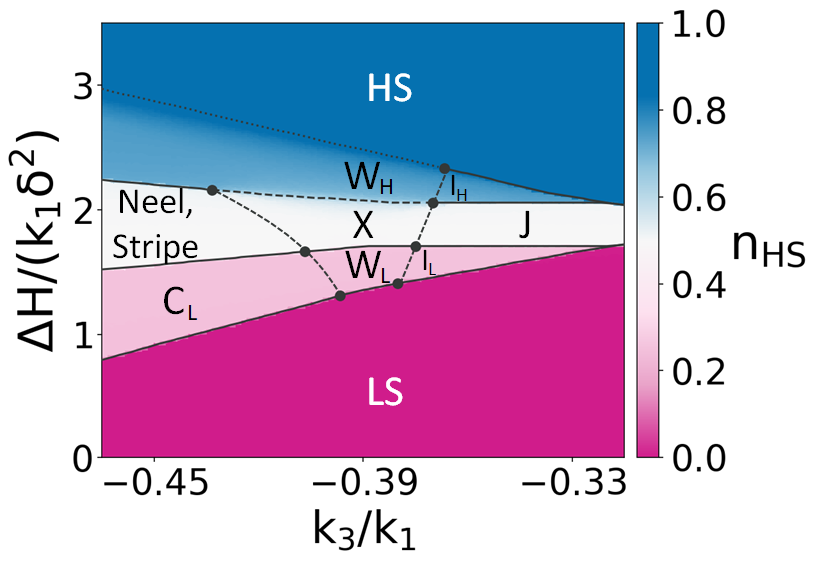}
	\caption{\label{fig:phaseDiafiniteT} Finite temperature phase diagram for $\Delta S=4\log 5$, $k_1>0$, $k_2=k_1/4$, $k_4=0$, and $k_5=0.0975k_1$. Shading represents the fraction of high spins, $n_{HS}\sim\chi T$, where $\chi$ is the magnetic susceptibility, calculated via parallel tempering. The solid black line represents the first order transitions, dashed black lines mark second order transitions, and the dotted line appears to be a crossover although could be a second order transition.
	}
\end{figure}

We plot a finite temperature slice of the phase diagram in \cref{fig:phaseDiafiniteT}. All of the spin-state crystal phases that are shown in the $T=0$ slice of the phase diagram (see \cref{fig:phaseDia0T}), occur at finite temperatures. This can be understood as a straightforward consequence of the single molecule entropy difference, $\Delta S$, which, following Wajnflasz and Pick,\cite{WP} we have absorbed into the Hamiltonian [Eq. (\ref{model})], which means that sweeping $T$ varies $\Delta G$. But, there is another contribution to the entropy -- the configurational entropy associated with the pattern of Ising pseduospins ($\{\sigma_i\}$).  This has a dramatic effect on the phase diagram, driving the emergence of   semectic W$_\text{L}$, W$_\text{H}$, and X phases, which are not found at zero temperature except precisely at first order lines.  In these phases we see precisely the same smectic order that we found in the  W$_\text{L}$, W$_\text{H}$, and X plateaus respectively (Figs. \ref{fig:nHS} and \ref{fig:snapshots}). 

\begin{figure}
	\includegraphics[width=\columnwidth]{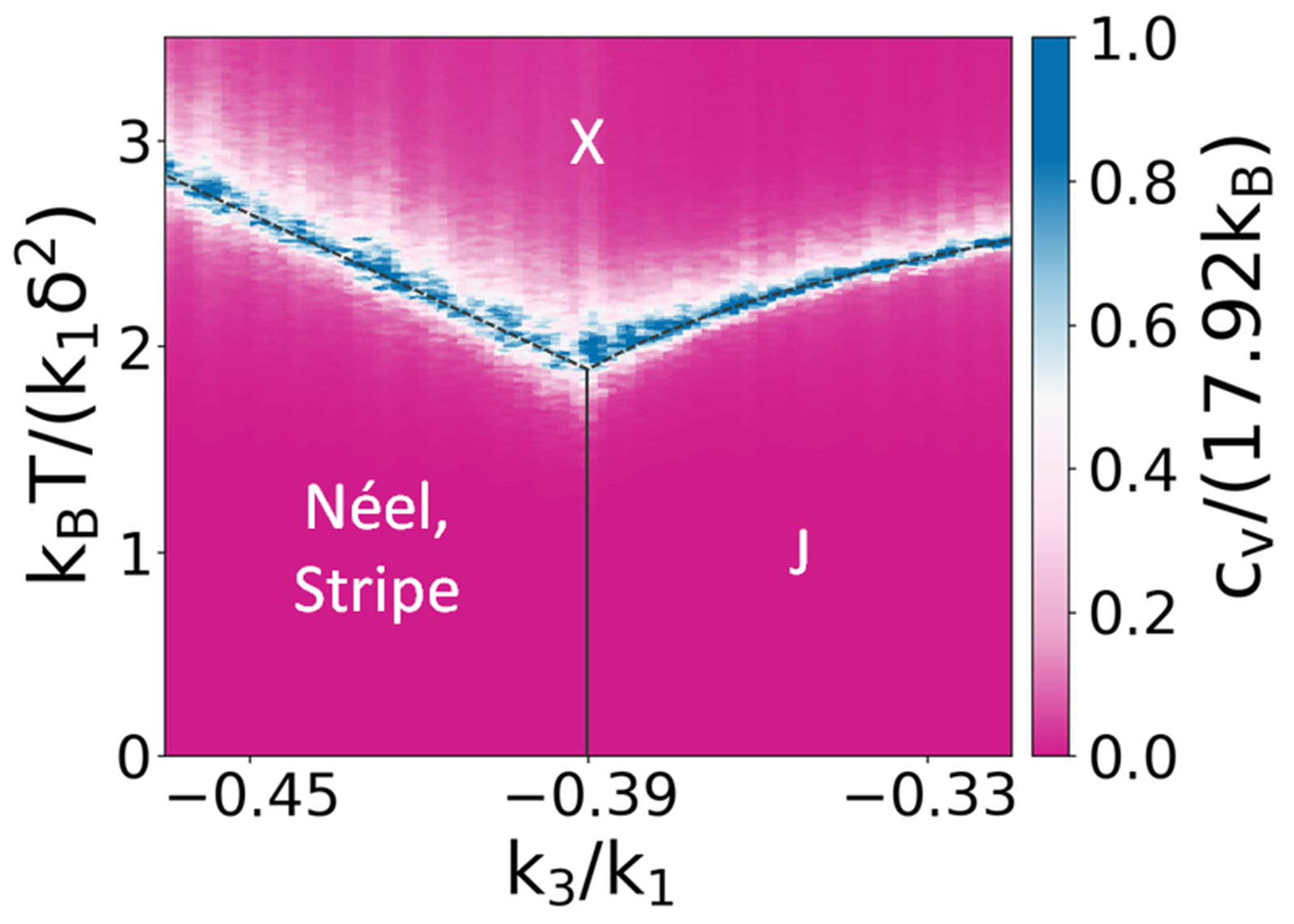}
	\caption{\label{fig:finiteTDS0} Finite temperature phase diagram  for $\Delta S=0$, $\Delta H=0$, $k_1>0$, $k_2=k_1/4$, $k_3=-0.39 k_1$, $k_4=0$, and $k_5=0.0975k_1$. Shading represents the calculated heat capacity, $c_V$, which has a maximum at the second order transition between the spin-state crystal phases and the spin-state smectic.  The solid black lines are first order transitions and dashed black lines (overlaying the blue shaded regions) are second order transitions.  
	}
\end{figure}

A simple demonstration that the configurational entropy plays a key role in stabilizing spin-state smectics can be given by setting $\Delta S=0$. This is no longer a realistic model of an SCO material, but nevertheless provides insight into the physics at play in them. In this limit varying the temperature does not change the free energy difference between HS and LS molecules ($\Delta G=\Delta H$) and the configurational entropy is the only game in town. Given a set of elastic interactions one can, by an appropriate choice of $\Delta H$, ensure that, for example, only  phases with, say, $n_{HS}=1/2$ occur as the temperature is swept. An example of this is shown in \cref{fig:finiteTDS0}. We find that the spin-state crystal phases at low temperatures  give way to a  spin-state smectics as the temperature is raised. %Like other liquid phases this can crossover to trivial uncorrelated (gas) phases as the temperature is further raised.

The smectic-X phase  occurs near a tricritcal point where the N\'eel, stripe and J  phases meet, \cref{fig:finiteTDS0}. This is important for understanding the origin of the smectic-X phase. We stressed above that the domain walls between the N\'eel and stripe  phases incur a significant energetic cost. %This is simple to confirm by microscopic calculation, but a more intuitive understanding can be gained by comparing Figs. \ref{fig:snapshots} and \ref{fig:sketch}.  
However, when these phases are also degenerate with the J  phase, the cost of forming a domain wall vanishes. This is because the J state order describes precisely the local configuration that is generated  by a domain wall between the N\'eel and stripe phases. 

It can be seen that in the snapshot of the smectic-X phase (\cref{fig:snapshots}) that every $3\times3$ square is short-range ordered in the same way as a $3\times3$ square in either the  N\'eel, stripe or J long-range ordered phases. As a $3\times3$ square contains all interactions up to fifth nearest neighbor, the dominant short range interactions cannot distinguish between the smectic-X phase and the three spin-state crystals along the first order line. However, the configurational entropy is larger for a spin-state smectic phase than for a spin-state crystalline phase -- it is essentially an entropy of mixing for the three long-range ordered phases. We have confirmed that this phase is robust to adding weak longer-range interactions. Therefore, at high enough temperatures the combined action of the configurational entropy and the strong short-range interactions favors the smectic-X phase. The smectic-X phase allows the system to increase the configurational entropy while only paying a small enthalpic penalty close to the tricritical point.

The constraint that every $3\times3$ square must display either N\'eel, stripe or J order in the smectic-X phase is analogous to the Bernal-Fowler ice rules\cite{Bernal}. It can be seen from the snapshot that this rule results in a state that is ordered in one direction (along the $x$-axis in \cref{fig:snapshots}) with alternating HS-LS-HS-LS-\dots. At first sight this is surprising as there is no long-range order in the one-dimensional Ising model for $T>0$.\cite{P&B} However, our findings are not inconsistent with this as we are dealing with a two dimensional system that is partially disordered. Thus, not only a $3\times3$ square that appears to display N\'eel, stripe or J order in the smectic-X phases but a $3\times L$ column, where $L$ is the linear dimension of the system. This state necessarily breaks rotational symmetry as only one direction can be long-ranged ordered. Thus, the competition between three crystalline spin-state orders naturally gives rise to long-range smectic-X order.

\begin{figure}	
	\includegraphics[width=0.9\columnwidth]{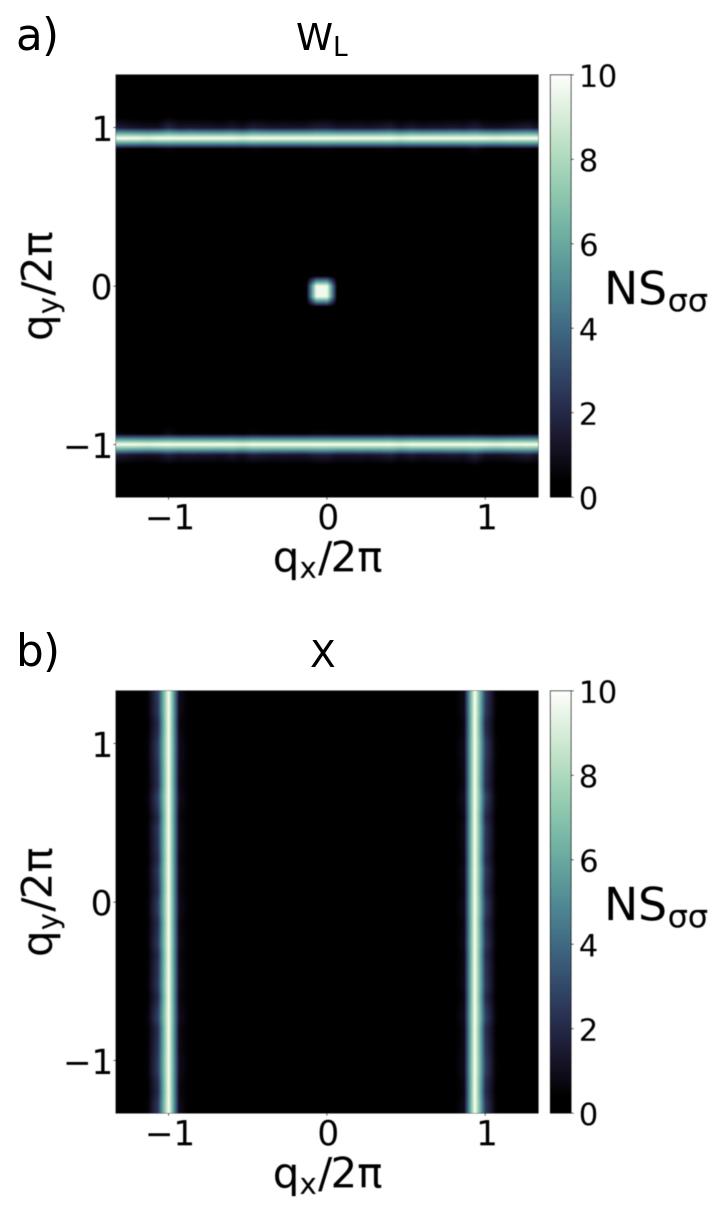}
	\caption{\label{fig:structurefactor} Pseudo-spin structure factors, $S_{\sigma\sigma}$ for the W$_\text{L}$ and X disordered antiferroelastic spin-states observed in our Monte Carlo simulations. Labels have the same meaning as \ref{fig:snapshots}. The structure factor for W$_{\text{L}}$ (X) shows distinct lines of Bragg peaks at $q_y\pm\pi$ ($q_x=\pm \pi$) indicating the existence of long-range order in one direction and disorder in the other. The direction which is ordered is not related to which phases occurs, but is a spontaneous breaking of rotational symmetry. The calculations were done at a) $\Delta H=4k_1\delta^2$ and b) $\Delta H=0$ with $\Delta S=k_B4\log(5)$, $k_BT=0.01$, $k_1>0$, $k_2=k_1/4$, $k_3=-0.39 k_1$, $k_4=0$ and $k_5=0.0975k_1$.}
\end{figure}

The W phases have an important difference from the X  phase -- they do not occur near tricritical points. Instead they occur near the ordinary critical points between the I and C phases. The I and C phases are both characterized by containing minority spin-states that are not nearest or next nearest neighbors --  neighboring minority spin-states are third nearest neighbors in the C phase and third and forth nearest neighbors in the I phase. Thus, domain walls between C phases displaced by one lattice constant from one another induce local configurations with I order and \textit{vice versa}. Therefore, the W phase only needs  two phases to be enthalpically competitive in order for the configurational entropy to stabilize a smectic phase. 

The W phases are specified by two rules: (a) no minority spin-state may have other minority spin-states as first or second nearest neighbors and (b) $n_{HS}=0.25$ (0.75) for W$_\text{L}$ (W$_\text{H}$). As with the X  phase this leads to long-range order in one direction (the $y$-direction in \cref{fig:snapshots}), it also leads to partial order in the perpendicular direction, with an alternating pattern of columns with equal numbers of both spin-states (ordered in 1D). However, there are only weak (and perhaps no) correlations between the ordering within neighboring columns. Thus,  the interplay competition between two competing crystalline spin-state orders gives rise to a long-range smectic-W order.

The spin-state ordering in the W and X phases gives rise to  lines of Bragg peaks in the pseudo-spin structure factor, $S_{\sigma\sigma}$. Depending on how rotational symmetry is broken  Bragg peaks occur along either $(q_x,q_y)=(\pm\pi,q)$ or $(q_x,q_y)=(q,\pm\pi)$, for arbitary $q$, providing direct evidence for smectic order. The featurelessness of the structure factor in one direction  indicates that there is little or no correlation between the antiferroelastically ordered columns of spin-states. The pseudo-spin structure factors can be directly mapped onto the spin scattering structure factor,\cite{JacePyro,JaceKagome} and is closely related to the positional structure factor  and is hence directly measurable by neutron or x-ray scattering experiments. 
 
The rules for the W and X phases allow us to make a Pauling-like estimate\cite{Pauling} of the residual entropy in these smectic phases. In the X phase there is long-range order in one direction, but in the perpendicular direction each may either be aligned or misaligned with its nearest neighbor (say the column to its left in \cref{fig:snapshots}). Thus, neglecting boundary effects, on an $N=L\times L$ lattice there are $2^L$ possible microstates, and $S_\text{residual}=L\,k_B\ln 2$. Similarly, in the W phases there is long-range order in one direction, but in the perpendicular direction each column containing both spin states may either be aligned or misaligned with its second nearest neighbor. Thus, neglecting boundary effects,  there are $2^{L/2}$ possible microstates, and $S_\text{residual}=\frac{L}{2}k_B\ln 2$. Thus, in the thermodynamic limit the specific entropy  ($s=S/N$) vanishes. This means that the W and X phases are less disordered than spin-state ices (which are concrete examples of spin-state liquid phases) where $s$ is a constant in the thermodynamic limit [$s_\text{Pauling}=(1/3)k_B\ln2$ on the kagome lattice and $k_B\ln(3/2)$ on the pyrocholre lattice],\cite{Contempt,JaceKagome,JaceSquare,Pauling} but more disordered than spin-state crystals where $s=0$. This quantifies the degree to which these spin-state smectic phases are intermediate between spin-state liquid and spin-state crystals.

\section{Conclusions}

In summary, we have shown that the competition between spin-state crystalline orders can give rise to partially disordered spin-state smectic phases in SCO materials. We have identified two  spin-state smectics in a well-known model of SCO materials.  Spin-state smectics display long-range spin-state order in one direction, but the spin-states are disordered in the perpendicular direction. Thus, the spin-state smectics are more ordered than spin-state liquids (such as spin-state ice), but less ordered than spin-state crystals (where there is long-range order in both directions, e.g., the phases shown in \cref{fig:sketch}). These different degrees of disorder are quantified by the residual entropies of the phases: the residual entropy of a spin-state crystal is independent of the system size; the residual entropy of a spin-state smectic scales with the linear dimension of the system; and the residual entropy of a spin-state ice scales with the total system size.

We showed that smectic phases give rise to plateaus in the fraction of HS molecules, and hence $\chi T$, similar to the intermediate plateaus that are caused by spin-state crystals. We have shown that spin-state smectics cause lines of Bragg peaks in the pseudo-spin structure factor, which should provide a definitive test for their existence. 

\begin{acknowledgments}
This work was supported by the Australian Research Council (grant no. DP200100305).
\end{acknowledgments}

\appendix

\bibliography{competing_orders}% Produces the bibliography via BibTeX.

\end{document}